\documentclass[aps,prb,twocolumn,groupedaddress,floatfix]{revtex4-2}

\usepackage{amsmath}
\usepackage{graphicx}
\usepackage{dcolumn}
\usepackage{bm}

\begin{document}

\title{Influence of Random Telegraph Noise on Quantum Bit Gate Operation }
\author{Jackson Likens$^1$, Sanjay Prabhakar$^1$, Ratan Lal$^2$ and Roderick Melnik$^2$}
\affiliation{
$^1$Department of Natural Science, D L Hubbard Center for Innovation, Loess Hill Research Center, Northwest Missouri State University, 800 University Drive, Maryville, MO 64468\\
$^2$Department of Computer Science, Northwest Missouri State University, 800 University Drive, Maryville, MO 64468\\
$^3$MS2Discovery Interdisciplinary Research Institute, M3AI Lab, Wilfrid Laurier University, 75 University Avenue, Waterloo, ON N3L 3V6, Canada
}
\date{Nov 3, 2022}

\begin{abstract}
We consider the problem of analyzing spin-flip qubit gate operation in presence of Random Telegraph Noise (RTN). Our broad approach is the following. We calculate the spin-flip probability of qubit driven by composite pulses, (Constant pulse (C-pulse), Quantum Well pulse (QW-pulse) and Barrier Potential pulse (BP-pulse)) in the presence of RTN using Feynman disentangling method. When composite pulses and RTN act in x-direction and z-direction respectively,  we calculate the optimal time to achieve $100\%$ spin-flip probability of qubit. We report the  shortcut of spin-flip qubit, which can be achieved by using C-pulse, followed by BP-pulse and QW-pulse. When jumps time in RTN are very fast, tuning of perfect fidelity or spin-flip probability extends to large RTN correlation time. On the other hand, when the jumps in RTN are very slow, the BP-pulse can be used to recover the lost fidelities.  Nevertheless, the fidelities of qubit gate operation   are larger than $90\%$, regardless of RTN jumps environments which may be beneficial  in quantum error correction. For more general case,  we have tested several pulse sequences for achieving high fidelity quantum gates, where we have used the pulses acting in different directions. From the calculations, we find high fidelity of qubit gate operation in presence of RTN is achieved when QW-pulse, BP-pulse and C-pulse act in x-direction, y-direction and z-direction,  respectively.
\end{abstract}



\maketitle


\section{Introduction}
Qubits can be manipulated in a desired fashion by excellent architect design in several physical devices, such as, quantum dots, cavity quantum electrodynamics, superconducting devices, Majorana fermions~\cite{choi00,burkard99,hu00,koh12,xiao10,prabhakar10,prabhakar14,prabhakar14a,
vool16,lau17,yoshihara17,clarke16,grunhaupt19,zhong19,kurpiers19,xu18,li2018perfect}. Manipulation of qubits in these devices seems promising in that one can make  quantum logic gates and memory devices for various quantum information processing applications. Such devices require sufficiently short gate operation time combined with long coherent time~\cite{loss98,golovach04,amasha08,balocchi11}.

When a qubit is operated on by a classical bit, then its decay time is given by a relaxation time which is also supposed to be longer than the minimum time required to execute one quantum gate operation. International Technology Roadmap for Semiconductors (ITRS) suggests that the node length and gate oxide thickness in CMOS technology for qubit gate operation is approaching approximately to one nanometer. Hence, a leakage current from source to drain through channel as well as gate to the channel through gate oxide layer is unavoidable. As a result recent  experimental observations in the oscillations of  drain current at both low and room temperatures confirms the origin of Random Telegraph Noise (RTN) that may reduce the performance of qubit gate operations~\cite{li18,liu18,singh18,zhan20,yang21}

\begin{figure*}
\includegraphics[width=18cm,height=5cm]{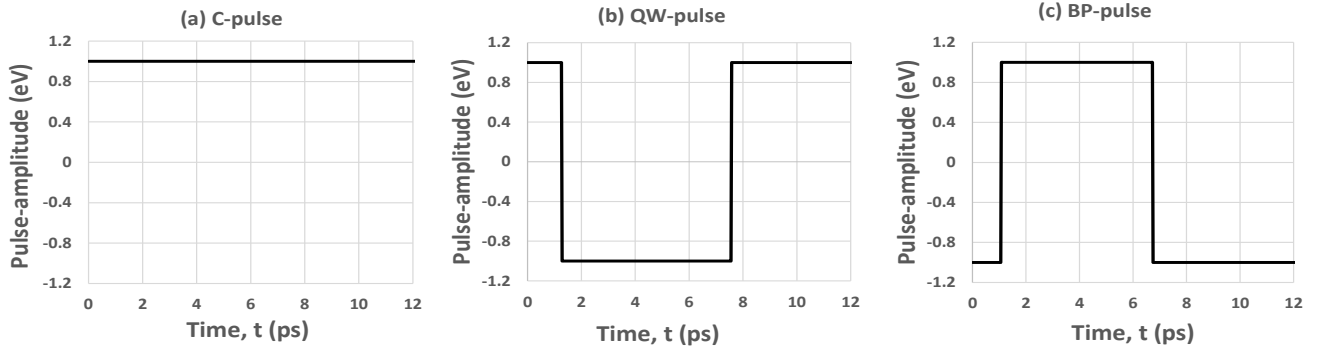}
\caption{\label{fig1} The designed pulses for (a) C-pulse, (b) QW-pulse, and (c) BP-pulse that operate on qubit to achieve high fidelity quantum gates under random telegraph noise. The functional form of these pulses are shown in Eqs.~(\ref{a-C}) and (\ref{a-BP/QW}). We chose $t_0 = 8$ps for C-pulse,  $t_0 =1.8$ps and $r_0=-0.6$ for BP-pulse and  $t_0 =2.0$ps and $r_0=2.56$ for QW-pulse.
}
\end{figure*}
\begin{figure*}
\includegraphics[width=15cm,height=8cm]{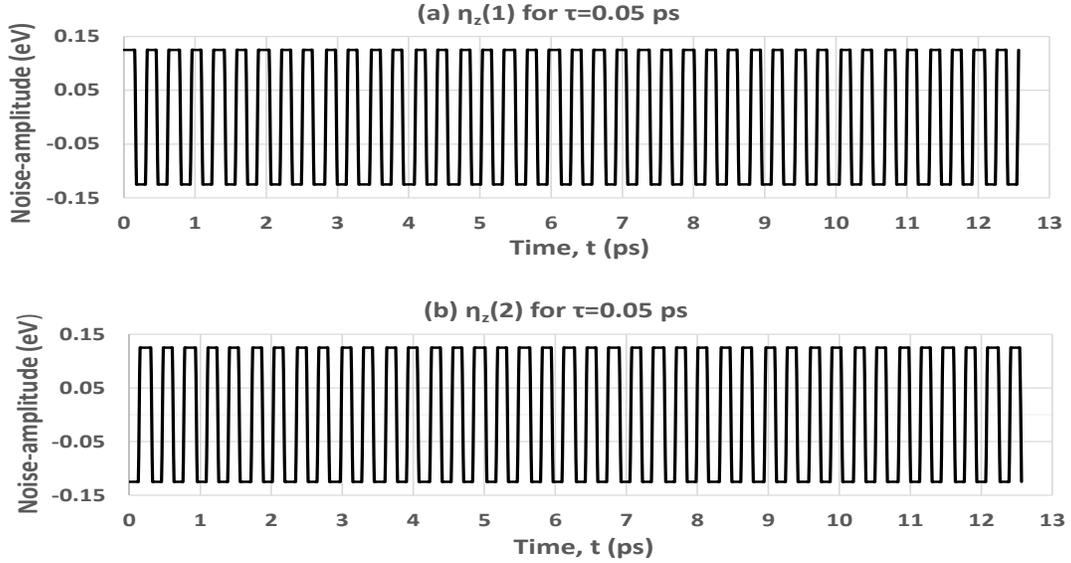}
\caption{\label{fig2} Simulations of random telegraph noise (RTN) as a function of time are obtained from Eq.~(\ref{eta-z}). Here we chose the correlation time $\tau_C=0.05$ps and $\Delta = 0.125$eV. Note that the density of RTN jumps between $\pm \Delta$ is random that is shown in Fig.\ref{fig2}(a) and (b). Here only two RTN functions are shown for demonstration purpose but in a realistic simulations of finding high fidelity of spin-flip quantum gates, 300 RTN trajectories have been chosen.)
}
\end{figure*}
\begin{figure}
\includegraphics[width=8.5cm,height=7cm]{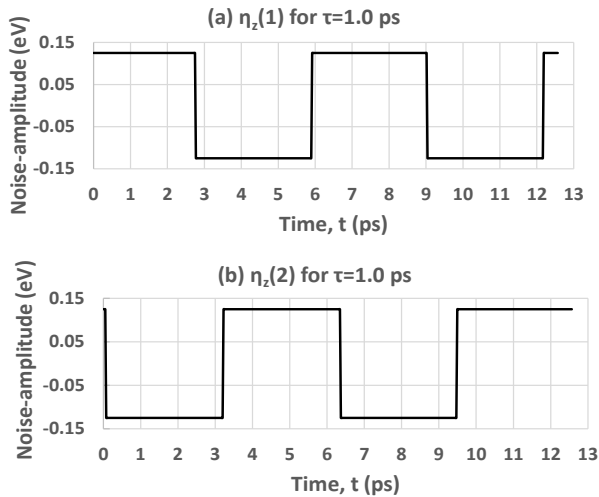}
\caption{\label{fig3} Same as to Fig.~\ref{fig2} but $\tau_c=1.0$ps. Notice that the jumps in RTN is significantly decreased as correlation time, $\tau_c$ increases from $0.05$ps in Fig.~\ref{fig2} to $1.0$ps in Fig.~\ref{fig3}. For large $\tau_c$ $\approx 5.0$ps, there are almost no jumps in RTN function and thus RTN can be treated as a white noise for large RTN correlation time.
}
\end{figure}
\begin{figure*}
\includegraphics[width=18cm,height=4.0cm]{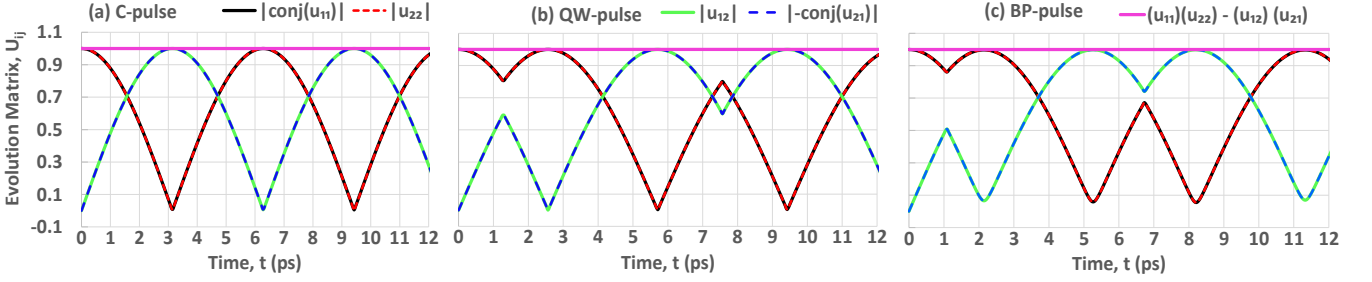}
\caption{\label{fig4} Components of the evolution operator (see Eq.~(\ref{Uk-7})) with respect to time for C-pulse in (a), QW-pulse in (b) and BP-pulse in (c). As can be seen in Figs.(a), (b) and (c), we find that  $|u_{11}^\star | = |u_{22}|$ and $|u_{12} | = |-u_{21}^\star |$ as well as $(u_{11}) (u_{22}) - (u_{12}) (u_{21}) = 1$. Hence, we confirmed that the components of the unitary time evolution operator obtained from Feynman disentangling operator techniques are in good agreement to the theoretical descriptions of the evolution matrix in quantum mechanics.
}
\end{figure*}
\begin{figure}
\includegraphics[width=8.5cm,height=4.5cm]{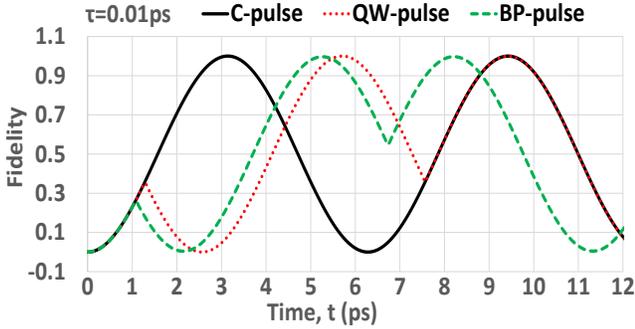}
\caption{\label{fig5} (color online) Fidelity of spin-flip qubit, obtained from Eq.~(\ref{phi}), as a function of RTN correlation time for C-pulse, QW-pulse and BP-pulse. Here, we chose $\Delta = 0.125$eV. As can be seen, tuning of perfect fidelity extends to large RTN correlation time for C-pulse due to short optimal gate operation time (t = 3.14 ps for C-pulse, t = 9.42 ps for QW-pulse and t = 8.20 ps for BP-pulse, see Fig.~\ref{fig5}). At large RTN correlation time, where there is no jumps in RTN, BP-pulse can be used to recovery the lost fidelities over other two pulses, (e.g. fidelity of BP-pulse is larger than C-pulse and QW-pulse at large RTN correlation time).
}
\end{figure}
\begin{figure}
\includegraphics[width=8.5cm,height=5.5cm]{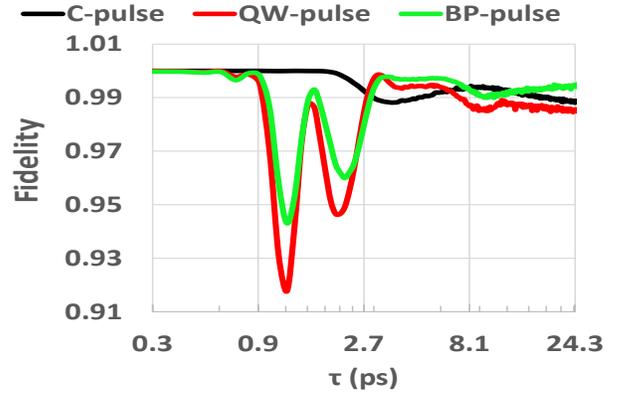}
\caption{\label{fig6} (color online) Fidelity of spin-flip qubit as a function of RTN correlation time, $\tau$ for C-pulse, QW-pulse and BP-pulse. We chose $\Delta = 0.125~eV$. As can be seen, C-pulse extends tuning of perfect fidelity to large RTN correlation time because it possesses very short optimal time, $t=3.14ps$ for spin-flip qubit gate operation. At large correlation time ($\tau\approx 20 ps$), recovery of fidelity for BP-pulse is  higher than all the other pulses.
}
\end{figure}

\begin{figure*}
\includegraphics[width=18.0cm,height=5.0cm]{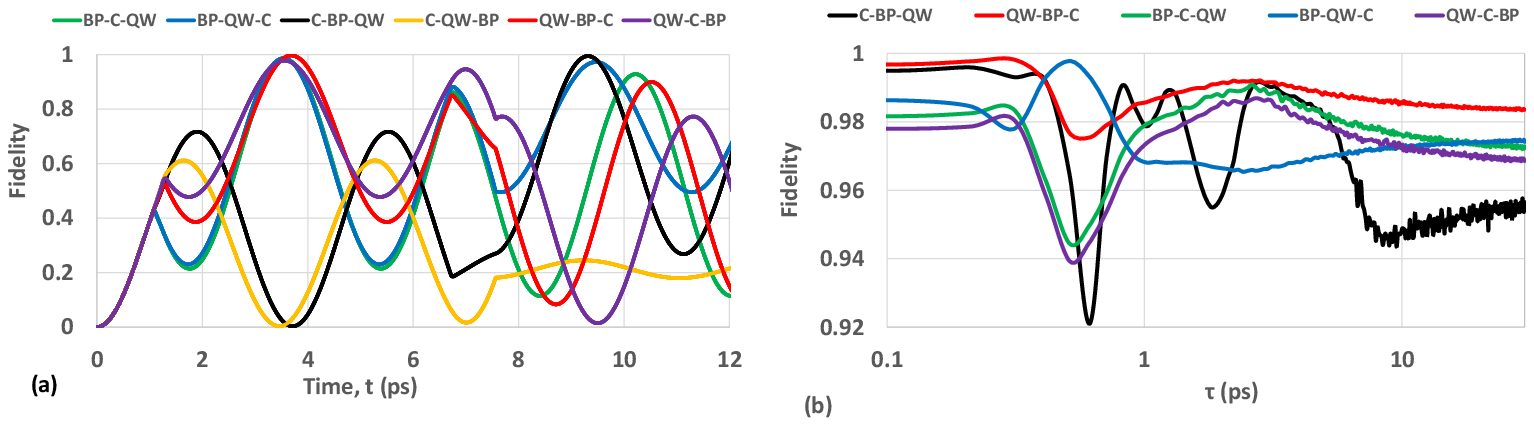}
\caption{\label{fig7} (a) Fidelity of spin-flip qubit as a function of time for the pulse sequences shown in the systematic orders, e.g. for BP-C-QW pulse sequence, BP-pulse acts in x-direction, C-pulse acts in y-direction and QW-pulse acts in z-direction. As can be seen the optimal time pulse for BP-C-QW is 3.57ps, for BP-QW-C is 3.53ps, C-BP-QW is 9.46ps, QW-BP-C is 3.66ps and QW-C-BP is 3.57ps. For the C-QW-BP pulse, fidelity is very small and may not be useful for achieving high fidelity quantum gates. (b) Fidelity of spin-flip qubit as a function of  RTN correlation time, $\tau$ for  $\Delta = 0.125~eV$ for several pulse sequences is shown. As can be seen QW-BP-C pulse has better performance than all the other pulses for achieving high fidelity quantum gates with respect to RTN correlation time.  Nevertheless the fidelities for all the pulses shown in Fig.~\ref{fig7}(b) is larger  than $92\%$. Such small error induced by RTN can be corrected  for application in qubit gate operation.
}
\end{figure*}

In most cases, compared to coherent time, the dephasing time of qubits in presence of noise is reduced by several orders of magnitude due to coupling of qubits to the environment. The reduction of dephasing time depends on the specific dynamical coupling sequence from where  the principle of quantum mechanics is inevitably lost.   Therefore, one might need to decouple the qubits from the environment and may consider more robust topological  method to preserve a quantum state, enabling robust quantum memory~\cite{june15,prabhakar10}
Hence, to make quantum computers, one needs to find an efficient and experimentally feasible algorithm that overcome  the issues  of undesired interactions of qubits with RTN or the environment because these interactions destroy the quantum coherence that lead to generate errors and loss of fidelity. In quantum computing language,  this phenomenon is called decoherence. For example, experimental observations reported that in GaAs quantum dots, decoherence time, $T_2^\star \approx 10 ns$ and coherent time, $T_1 \approx 0.1 ms$, whereas  for Si, $T_2^\star \approx 100 ns$ and $T_1 \approx 0.1 ms$~\cite{bluhm11,petta05,maune12,pla12,malinowski18,friesen17,martins17,he19,geng16,liu19,throckmorton17,yang18neural,huang17}. There are several possible ways to overcome the issues of decoherence, as for example, fidelity recovery by applying error-correcting codes, decoherence free subspace coding, noiseless subsystem coding, dynamical decoupling from hot bath, numerical design of pulse sequences, that is  more robust to experimental
inhomogeneities, and optimal control pulse based on Markovian
master equation descriptions~\cite{ofek16,michael16,brown16,reiserer16,lau16,liu16,castro12,wang12,jenei19,rudge19,motzoi16,dickel18,kurpiers18,campagne18,dridi20}.

In this paper, we design several control pulses acting on
a single bit-flip computational basis states in presence of Random Telegraph Noise (RTN). Note that the choice of modeling parameters for RTN is same as  that of experimentally observed RTN in Ref~\cite{li18}. For example, choosing small noise correlation time in RTN provides very fast jumps, less than one tenth of picosecond of amplitude in RTN, whereas large noise correlation time provides slow jumps, larger than about 2 picosecond in RTN. Hence, checking performance of qubit under RTN in this paper resembles the realistic form of RTN that recently discussed in experimental studies in Ref.~\cite{li18}. The present work identifies  different regimes of operating parameters in the designed control pulses that eliminate the series of phase and dynamical errors and increase the recovery of  high fidelities of spin-flip qubit gate operation. The designed composite pulses, that we named, are Constant pulse (C-pulse), Quantum Well pulse (QW-pulse) and Barrier Potential pulse (BP-pulse), eventually acting on a qubit in presence of  Random Telegraph Noise (RTN). The amplitude of C-pulse is constant with time, whereas three composite pulse sequences of different time width form QW-pulse and BP-pulse. The calculations of spin-flip qubit gate operation under RTN at various noise correlation times as well as various energy amplitudes of noise strength provide an indication of most efficient way to perform algorithm for achieving high fidelity quantum gates for quantum circuits and quantum error correction. In this paper, we report that when the qubits are driven by pulses in the x-direction and the RTN act in the z-direction then the C-pulse induce less systematic errors over QW-pulse and BP-pulse. For a more general case, we have tested all the possible combinations of the pulses acting in arbitrary x, y and z directions in presence of RTN and show that maximum fidelity of qubit gate operation can be achieved if C-pulse acts in x-direction, BP-pulse acts in y-direction and QW-pulse acts in z-direction. This useful information may be utilized to identify experimentally feasible pulses in presence of RTN for the design of next generation quantum circuits.

The paper is organized as follows. In section~\ref{theoretical-model}, we provide a theoretical description of finding exact unitary operator using Feynman Disentangling operator scheme of  the model Hamiltonian of a qubit driven by several control pulses in presence of  RTN. In section~\ref{results-discussions}, we analyze two main results: (i) fidelity of qubits driven by a pulse in the x-direction and RTN in z-direction. and (ii) the fidelity of qubits driven by individual pulses that act in the x,y and z-direction and the RTN still in the z-direction. Finally we conclude the results in Section~\ref{conclusion}.

\section{Model Hamiltonian}\label{theoretical-model}
The Hamiltonian of a single qubit is written as~\cite{nelson}
\begin{equation}
H(t)=\sum_{i\in \{x,y,z\}} \frac{1}{2}\left[a_i(t)+ \eta_i(t)\right]\cdot \sigma_i,
\label{Ht}
\end{equation}
where $a_i(t)$ is the energy amplitude of the external control pulse,  $\eta_i(t)$ is the energy amplitude of the Random Telegraph Noise and $\sigma_i$ is the Pauli spin matrices. Our goal is to design several composite pulses that provide high probability of spin-flip qubit in presence of RTN. Hence, we model the mathematical function of pulses as:
\begin{equation}
a_C(t)= \frac{\cos(t/t_0)}{\left|\cos(t/t_0)\right|},
\label{a-C}
\end{equation}
where $t_0=8$ps for C-pulse. For BP  and QW composite pulses, we model the function as:
\begin{equation}
a_{BP/QW}(t)= \frac{\sin(t/t_0 + r_0)}{\left|\sin(t/t_0 + r_0)\right|},
\label{a-BP/QW}
\end{equation}
where $t_0=1.8$ps and $r_0=-0.6$ for BP-pulse and $t_0=2.0$ps and $r_0=2.56$ for QW-pulse. The designed C-pulse, BP-pulse and QW-pulse obtained from Eqs.~(\ref{a-C}) and (\ref{a-BP/QW}) are shown in Fig.~\ref{fig1}. Notice that amplitude of C-pulse is constant whereas combination of three composite pulses that form QW-pulse is shown in Fig.~\ref{fig1}(b) and the combination of three composite pulses that form BP-pulse sequence is shown in Fig.~\ref{fig1}(c). In the  Hamiltonian~(\ref{Ht}), the RTN only acts in the z-direction  because  random RTN jumps originated mostly due to leakage current from gate oxide to the channel as spins are transported from source to  drain, as demonstrated experimentally in~\cite{li18}. The energy amplitude of the RTN changes randomly between $-\Delta$ and $\Delta$, where $\Delta$ is the maximum energy amplitude. Hence, we model the RTN trajectory as:
\begin{equation}
\eta_{z}(t)= \Delta \frac{\sin(t/\tau  - r_d)}{\left|\sin(t/\tau - r_d)\right|},
\label{eta-z}
\end{equation}
where $\tau$ is RTN correlation time and $r_d = \log(p_i),$ where $p_i \epsilon (0,1)$ is the random numbers that determine the probability of  random jumps of RTN trajectories.
Two RTN functions are shown in Fig.~\ref{fig2} and \ref{fig3}. For the realistic simulations of spin-flip qubit, we have chosen $300$ randomly generated RTN functions. As can be seen in Fig.~\ref{fig2}, there are large density of RTN jumps in the vicinity of zero correlation times, $\tau_c$. On the other hand, as $\tau_c$ increases, the density of RTN jumps decreases that can be seen in Fig.~\ref{fig3}. Note that modeling parameters of RTN  trajectories shown in Figs.~\ref{fig2} and \ref{fig3} are in close vicinity  of experimental trajectories of RTN reported in~\cite{li18}. To find the system dynamics,  average over different RTN sample trajectories is chosen to find the density matrix:
\begin{equation}
\rho(t)=   \lim_{N\to\infty}   \frac{1}{N} \sum_{k=1}^N U_k(t) \rho_0 U_k^\dagger(t),
\label{rho}
\end{equation}
where $\rho_0$ is the initial state of the system and $\{U_k \}$ is the unitary time evolution of the qubit under the influence of control pulses (see Fig.~\ref{fig1}) and RTN (see Figs.~\ref{fig2} and \ref{fig3}). We write the unitary time evolution operator as
\begin{equation}
U_k(t)= T  e^{{-i/\hbar}\int_0^t dt H(t) },
\label{Uk}
\end{equation}
where $T$ is the time ordering parameter.  We apply Feynman disentagling operators scheme and find the evolution operator as follows. The Hamiltonian, $H(t)$ in (\ref{Ht}) or (\ref{Uk}) can be written as
\begin{equation}
H(t) = H_+s_+ + H_-s_- + H_zs_z,
\label{Hta}
\end{equation}
where,
\begin{equation}
H_+ = \frac{1}{2} \left(a_x(t) - i a_y (t) \right)
\label{H+}
\end{equation}
\begin{equation}
H_- = \frac{1}{2} \left(a_x(t) + i a_y (t) \right)
\label{H-}
\end{equation}
\begin{equation}
H_z =  a_z(t) + \eta_z(t),
\label{Hz}
\end{equation}
and $s_{\pm}=\left(\sigma_x \pm i\sigma_y\right)/2$ and $s_z = \sigma_z/2$. In the disentangled form, the unitary time evolution operator~(\ref{Uk}) can be written as
\begin{equation}
U_k(t)= \exp(\alpha (t) s_+) \exp(\beta(t) s_z) \exp(\gamma (t) s_-),
\label{Uk-1}
\end{equation}
where $\alpha(t)$, $\beta(t)$ and $\gamma(t)$ are unknown  that can be found by using Feynman disentangling operator calculus method~\cite{feynman51,popov07,prabhakar10,prabhakar14a}. We write $H(t)$ of (\ref{Hta}) as
\begin{equation}
H(t) = \xi s_+^\prime + (H_+ -\xi)s_+^\prime +H_z s_z^\prime +H_- s_-^\prime,
\label{Hta-1}
\end{equation}
where
\begin{equation}
\alpha(t) = -\frac{i}{\hbar} \int_0^t \xi (t) dt,
\label{alpha}
\end{equation}
\begin{equation}
s_\mu^\prime (t) = \exp(-\alpha s_+) s_\mu \exp(\alpha s_+).
\label{s-mu}
\end{equation}
Differentiating Eq.~(\ref{s-mu}) with respect to $\alpha$, we can write
\begin{equation}
\frac{ds_\mu^\prime (t)}{d\alpha} = \exp(-\alpha s_+) \left[ s_\mu, s_+ \right]s_\mu \exp(\alpha s_+),
\label{ds-mu-1}
\end{equation}
and utilizing initial condition, $s_\mu^\prime (0) =s_\mu$, we find $s_+^\prime = s_+$, $s_0^\prime = s_0 +  s_+ \alpha $, $s_-^\prime = s_- - s_+ \alpha ^2 -2 s_0 \alpha$.
Substituting Eq.~(\ref{Hta-1}) in Eq.~(\ref{Uk}), we can write the unitary time evolution operator as
\begin{equation}
U_k(t)= e^{\alpha (t) s_+} T e^{-\frac{i}{\hbar} h(\alpha) dt },
\label{Uk-2}
\end{equation}
where
\begin{equation}
\begin{split}
h(\alpha) = & \left(H_+ -\xi +H_z \alpha - H_- \alpha ^2  \right) s_+  + H_z s_z  \\
            & + H_- s_- - 2 H_- s_z \alpha.
\end{split}
\label{h-alpha}
\end{equation}
Equating coefficient of $s_+$ to zero, we write
\begin{equation}
\frac{d\alpha }{dt} = -\frac{i}{\hbar} \left[\frac{1}{2} \left(a_x - i a_y\right)  + \left(a_z +\eta_z \right)\alpha - \frac{1}{2} \left(a_x + i a_y\right)\alpha ^2 \right].
\label{dalpha}
\end{equation}
Hence, $s_+$ of (\ref{h-alpha}) in (\ref{Uk-2}) is completely disentangled and thus unitary time evolution operator~((\ref{Uk-2})) can be written in the disentangled form as
\begin{equation}
U_k(t)= e^{\alpha (t) s_+} T e^{-\frac{i}{\hbar}\int_{0}^{t} H^\prime \left( s^\prime_\mu, \alpha\right) dt},
\label{Uk-3}
\end{equation}
where,
\begin{equation}
H^\prime (s_\mu^\prime, \alpha) = \varsigma s_z^\prime +  \left(H_z - 2 H_- \alpha -\varsigma\right) s_z^\prime   +H_- s_-^\prime .
\label{H-prime-s}
\end{equation}
\begin{equation}
\beta(t) = -\frac{i}{\hbar} \int_0^t \varsigma (t) dt,
\label{beta}
\end{equation}
\begin{equation}
s_\mu^\prime (t) = \exp(-\alpha s_z) s_\mu \exp(\alpha s_z).
\label{s-mu-1}
\end{equation}
Differentiating Eq.~(\ref{s-mu-1}) with respect to $\beta$, we can write
\begin{equation}
\frac{ds_\mu^\prime (t)}{d\alpha} = \exp(-\alpha s_z) \left[ s_\mu, s_z \right]s_\mu \exp(\alpha s_z),
\label{ds-mu-1}
\end{equation}
and utilizing initial condition, $s_\mu^\prime (0) =s_\mu$, we find $s_z^\prime = s_z$, $s_-^\prime = s_- \exp (\beta)$.
Substituting Eq.~(\ref{H-prime-s}) in Eq.~(\ref{Uk-3}), we can write the unitary time evolution operator as
\begin{equation}
U_k(t)= e^{\alpha (t) s_+} e^{\beta (t) s_z}
T e^{-\frac{i}{\hbar} \left[ \left(H_0 - 2H_- \alpha - H_- \varsigma  \right) s_z + H_- s_- \right]dt }.
\label{Uk-4}
\end{equation}
Equating coefficient of $s_z$ to zero, we write
\begin{equation}
\frac{d\beta }{dt} = -\frac{i}{\hbar} \left[ a_z +\eta_z - \left(a_x + ia_y\right) \right].
\label{dbeta}
\end{equation}
Hence, $s_z$ of~(\ref{Uk-4}) is completely disentangled and thus unitary time evolution operator \ref{Uk-4} can be written in the disentangled form as
\begin{equation}
U_k(t)= e^{\alpha (t) s_+} e^{\beta (t) s_z} T e^{-\frac{i}{\hbar}\int_{0}^{t} H^{\prime\prime} \left( s^\prime_\mu, \alpha,\beta\right) dt},
\label{Uk-5}
\end{equation}
where,
\begin{equation}
H^{\prime\prime} (s_\mu^\prime, \alpha, \beta) = \chi s_-^\prime +  \left(H_- e^{\beta} - \chi \right) s_-^\prime   .
\label{H-prime-s1}
\end{equation}
\begin{equation}
\gamma(t) = -\frac{i}{\hbar} \int_0^t \chi (t) dt,
\label{gamma}
\end{equation}
\begin{equation}
s_\mu^\prime (t) = \exp(-\gamma s_-) s_\mu \exp(\alpha s_-).
\label{s-mu-2}
\end{equation}
Differentiating Eq.~(\ref{s-mu-2}) with respect to $\gamma$, we can write
\begin{equation}
\frac{ds_\mu^\prime (t)}{d\gamma} = \exp(-\gamma s_-) \left[ s_\mu, s_- \right]s_\mu \exp(\alpha s_-),
\label{ds-mu-3}
\end{equation}
and utilizing initial condition, $s_\mu^\prime (0) =s_\mu$, we find $s_-^\prime = s_-$.
Substituting Eq.~(\ref{H-prime-s1}) in Eq.~(\ref{Uk-4}), we can write the unitary time evolution operator as
\begin{equation}
U_k(t)= e^{\alpha (t) s_+} e^{\beta (t) s_z} e^{\gamma (t) s_-}
T e^{-\frac{i}{\hbar} \left[ \left( H_-\exp(\beta) - \chi \right)s^\prime_-\right]dt }.
\label{Uk-6}
\end{equation}
Equating coefficient of $s_z$ to zero, we write
\begin{equation}
\frac{d\gamma }{dt} = -\frac{i}{\hbar} \left[\frac{1}{2} \left(a_x + i a_y\right)\right].
\label{dgamma}
\end{equation}
Hence, unitary time evolution operator (\ref{Uk-6}) is completely disentangled. Finally, the exact unitary time evolution operator (\ref{Uk-6}) can be written as
\begin{eqnarray}
U_k(t)=\left(\begin{array}{cc}\exp\left({\frac{\beta}{2}}\right) + \alpha \gamma \exp\left\{{-\frac{\beta}{2}}\right\} & \alpha\exp\left(-{\frac{\beta}{2}}\right)\\
\gamma\exp\left(-{\frac{\beta}{2}}\right) & \exp\left(-{\frac{\beta}{2}}\right) \end{array}\right),~~~~~ \label{Uk-7}
\end{eqnarray}
Since (\ref{Uk-7}) is the exact time evolution operator in the disentangled form, we can construct the density matrix of (\ref{rho}) and find the fidelity of qubit as
\begin{equation}
\phi= tr\{\rho_f^\dagger \rho_T \},
\label{phi}
\end{equation}
where $\rho_f=U_f \rho_0 U_f^\dagger$ is the final state of the system and  $\rho_T$ is the final desired state of the qubit, where the pulse sequence end. Here $U_f$ is a quantum gate that is independent of the initial state preparation. In this paper, we chose $U_f$ is a Pauli X-gate and prepare the initial state $|0> = \left(0, 0; 0, 1\right)_{2\times 2}$ and final state $|1>=\left(1, 0; 0, 0\right)_{2\times 2}$ and then find the spin-flip probability by using Eq.~(\ref{phi}). In this paper,  we use Feynaman disentangling operator scheme~\cite{feynman51,popov07,prabhakar10,prabhakar14a} to find the system dynamics of  Eq.~(\ref{Uk}). We have chosen $300$ RTN trajectories and $\hbar = a_{max} = 1$.

\section{Results and Discussions}\label{results-discussions}

In Fig.~\ref{fig4}, we have plotted the components of the evolution operator~(\ref{Uk-7}) of C-pulse in (a), QW-pulse in (b) and BP-pulse in (c) for RTN correlation time, $\tau_C =0.001 ps$. The data in these  plots show that $u_{22} = |conj(u_{11})|$ and $u_{12} = |-conj(u_{21})|$ for C-pulse, QW-pulse and BP-pulse as well as $|U_k\left(t,0\right)|=1$. Hence, we confirmed that the evolution matrix obtained from Feynman disentangling operator method is very accurate. In Fig.~\ref{fig5}, we have plotted the fidelity of spin-flip qubit with respect to the evolution of time for C-pulse, QW-pulse and BP-pulse. As can be see that the perfect fidelity can be achieved at $t=3.14$ps for C-pulse, $t=9.42$ps for QW-pulse and $t=8.20$ps for BP-pulse. We consider these times as the optimum time for achieving high fidelity of spin-flip qubit in presence of RTN. Notice that optimum time for spin-flip qubit is smallest for C-pulse (i.e., $t=3.14$ps) but optimum time for BP-pulse is smaller (i.e., $t=8.20$ps) than the QW-pulse (i.e., $t=9.42$ps) because C-pulse has constant energy amplitude but width of the QW-pulse is larger than the width of BP-pulse (e.g., see Fig.~\ref{fig1}). In other words, QW-pulse lasts longer than the BP-pulse during the spin-flip qubit gate operation.

We consider the C-pulse, QW-pulse and BP-pulse act in the x-direction and RTN in z-direction in the Hamiltonian (\ref{Ht}) and plotted the fidelity of spin-flip qubit with respect to the RTN corellation time in Fig.~\ref{fig6}. As can be seen in the regime of vanishing RTN correlation time ($i.e., \tau_c \rightarrow 0$), perfect fidelity can be observed for all the three pulses, e.g.,  C-pulse, QW-pulse and BP-pulse due to the fact that the single qubit   does not have sufficiently large time to drift along the direction of  densely populated random telegraph noise (see Fig.~\ref{fig2}). Note that the RTN function is very dense in the vicinity of zero correlation time (see Fig.\ref{fig2}) but has no jumps in the vicinity of infinite correlation time (e.g., density of noise jumps decreases as $\tau_c$ increases, see Fig.\ref{fig2} and Fig.\ref{fig3}). In Fig.~\ref{fig6}, we also observed oscillations in the intermediate regime of RTN correlation time, (e.g., $\tau \approx 2.7$ps for C-pulse and $\tau \approx 1.2$ps for BP-pulse and QW-pulse) due to the fact that as the jumps in RTN slowed down then qubit stays in the RTN state that let the quibit to drift along the noise direction. For large RTN correlation time where RTN can be treated as a white noise, or there is no jumps in RTN,  qubit recovers most of its lost fidelity. As can be seen, when RTN is treated as a white noise, BP-pulse has high fidelity than C-pulse and QW-pulse.

As shown in Fig.~\ref{fig7}, for the pulses acting in all three directions and RTN in z-direction, we can achieved high fidelity of spin-flip qubit only when QW-pulse acts in x-direction, BP-pulse acts in y-direction and C-pulse acts in z-direction (red plot in Fig.~\ref{fig7}(b)). Optimum time for such pulse sequences is $t=3.66$ps that can be achieved from fidelity of spin-flip qubit-vs-time in Fig.~\ref{fig7} (a). Note that we have tested all the other possible combinations of pulses as shown in Fig.~\ref{fig7}(a) to achieve the optimal time for high fidelity of qubit. In Fig.~\ref{fig7}(b), we have plotted the fidelity of spin-flip qubit with respect to the RTN correlation time. As can be seen, the fidelity of QW-BP-C pulse in presence of RTN has better performance than all the other sequences of the pulses but the fidelity of spin-flip qubit for all the other pulses are still above $90\%$ and may still  be used for quantum error correction and quantum information processing.

\section{Conclusion}\label{conclusion}

We have shown a possible way to achieve high fidelity of spin-flip qubit gate operation using several control pulses (e.g., C-pulse, QW-pulse, and BP-pulse) in presence of random telegraph noises.  In Fig.~\ref{fig6}, in the limit of vanishing RTN correlation time, C-pulse can be used for the measurement of achieving high fidelity spin-flip qubit operation due to its small optimal gate operation time. In particular, for small values of RTN correlation time, C-pulse extends the tuning of perfect fidelity to large RTN correlation time.  When the RTN correlation time is large (e.g., $\tau_C \rightarrow \inf $), possibly white noise, BP-pulse can be used to achieve high fidelity of spin-flip qubit gate operation. In Fig.~\ref{fig7} when
the pulses acting in all the three directions, we have tested several pulse sequences for achieving high fidelity quantum gates and report that QW-pulse acting in x-direction, BP-pulse acting in y-direction
and C-pulse acting in z-direction can be used to provide high fidelity of qubit gate operation in presence of RTN.  Regardless of RTN conditions, the fidelities of spin-flip qubit gate operations for  these pulses are larger than $92\%$. Since the modeling parameters of RTN resembles with the experimentally observed RTN in Ref.~\cite{li18}, one can use QW-pulse, BP-pulse and C-pulse to transport the qubit for several different kinds of quantum gate operation that may have applications in solid state realization of quantum computing and quantum information processing.

\section{Acknowledgements}
The simulations are performed at BARTIK High-Performance Cluster (National Science Foundation Grant No. CNS-1624416, U.S.) in Northwest Missouri State University. JL and SP acknowledge Northwest Missouri State University for providing financial support to present these results at American Physical Society March meeting in Chicago (2022). RM is acknowledging the
support of NSERC Discovery and CRC Programs.

%

\end{document}